\documentclass{eas}
\usepackage{graphicx}
\begin{document}

\title{Age Dependence of the Vertical Distribution of Young Open Clusters:
Implications for Local Mass Density, Stellar Evolution, and Star
Formation}
\author{A.~K.~Dambis}\address{Sternberg Astronomical Institute, Universitetskii pr. 13,
Moscow, 119992 Russia}
%
%

\begin{abstract}
The ages of 203 open clusters from the list of A.K.Dambis
(\cite{dambis}) are computed in terms of Cambridge evolutionary
tracks with and without the allowance for convective overshooting
(Pols {\em et al.\/} \cite{pols}). The vertical scaleheight of the
cluster layer for 123 objects at Galactocentric distances
$R_0$--1 kpc $< R_g< R_0+$ 1 kpc is found to vary
nonmonotonically with age exhibiting a wavelike pattern similar
to the one earlier found for the Cepheid population (Joeveer
\cite{joeveer}). The period of these variations is equal to $P_Z$
= 74 $\pm$ 2 Myr and $P_Z$ = 92 $\pm$ 2 Myr if cluster ages are
computed in terms of evolutionary models of Pols {\em et al.\/}
(\cite{pols}) without and with overshooting, respectively. If
interpreted as a manifestation of vertical virial oscillations,
the implications of the pattern found are threefold: (1) the
period of vertical oscillations can be reconciled with the known
local density of visible matter only if cluster ages are computed
with no or just mild overshooting ($P_Z$ = 74 $\pm$ 2 Myr implies
a maximum local density of $\rho$ = 0.118 $\pm$ 0.006 $M_{\odot}$
pc$^{-3}$ compared to $\rho$ = 0.102 $M_{\odot}$ pc$^{-3}$
recently inferred from Hipparcos data (Holmberg \& Flynn
\cite{hf}), whereas the period implied by the ages computed using
models with overshooting ($P_Z$ = 92 $\pm$ 2 Myr)  implies a
maximum local density of only $\rho$ = 0.075 $\pm$ 0.003
$M_{\odot}$ pc$^{-3}$ and is thus totally incompatible with recent
estimates, (2) there is no much room left for the dark matter
($\rho_{DM} \le$ 0.027 $M_{\odot}$ pc$^{-3}$ in the Galactic disk
near the solar Galactocentric distance, and (3) at the time of
their formation open clusters have, on the average, excess
kinetic energy (in the vertical direction) and as a population
are not in virial equilibrium; moreover, the initial vertical
coordinates of open clusters (at the time of their birth) are
strongly and positively correlated with initial vertical
velocities (the correlation coefficient is $r(Z_0,V_{Z(0)})$ =
0.81 $\pm$ 0.08), thus favoring a scenario where star formation in
the disk is triggered by some massive objects falling onto the
Galactic plane.
\end{abstract}
\maketitle
\section{Introduction}
Almost 30 years ago Joeveer (\cite{joeveer}) found the dispersion
of vertical coordinates ($\sigma Z$) of Galactic classical
Cepheids to vary nonmonotonically with age (inferred from period)
in a wavelike pattern. He interpreted these oscillations as a
manifestation of the fact that each subpopulation of coeval
Cepheids is not in vertical virial equilibrium, i.e., that the
mean initial (vertical) velocities and coordinates of stars are
not perfectly balanced. This causes the stars at a certain
Galactocentric distance to recede from and approach the Galactic
plane in a correlated way with a period (frequency) determined by
local mass density (and therefore the same for all stars at a
given distance from the Galactic center). Joeveer
(\cite{joeveer}) used the period of vertical oscillations of
Galactic Cepheids thus determined to estimate the local mass
density in the solar neighborhood.

Here we analyze the evolution of the vertical distribution of
another class of young Galactic-disk objects with accurately
determinable ages - - open star clusters. However, unlike Joeveer
(\cite{joeveer}), we consider our primary task to be not to
determine local mass density, which we believe to be much better
constrained by recent kinematic analyses and therefore already
known, but to use this already known density to discriminate
between two evolutionary grids (Pols {\em et al.\/} \cite{pols})
that are identical in all respects except that one is computed
with and another without the allowance for convective
overshooting (these two grids, naturally,  yield different
cluster ages and, consequently, different periods of vertical
oscillations of the cluster population).

\section{Basic formulas}

Consider an open cluster located in the vicinity of the Galactic
plane. If the cluster has a close-to-circular velocity in the
Galactic plane and small vertical velocity, its vertical motion
is, to a good approximation, decoupled from the motion parallel
to the Galactic plane and obeys the following equation:
\begin{equation}
d Z^2/dt^2 + \omega_z^2 \cdot Z = 0
\end{equation}
(see, e.g., King \cite{king}), where $\omega_z$ (=2$\pi/P_z$) is
the frequency (and $P_z$, the period) of vertical oscillations
determined by local mass density and rotation- curve parameters
via the Poisson equation:
\begin{equation}
(\omega_z)^2+(d^2\Phi/dR^2)+(1/R)(d\Phi/dR) = 4\pi G\rho.
\end{equation}
The sum of the last two terms in the left-hand side of the above
equation can be easily expressed in terms of the local values of
Oort's constants $A$ and $B$:
\begin{equation}
(d^2\Phi/dR^2)+(1/R)(d\Phi/dR) = -2\cdot (A^2 - B^2)
\end{equation}
and shown to be negligible compared to $4\pi G \rho$. Young
objects are close to the Galactic plane and therefore for clusters
located within a narrow interval of Galactocentric distances
$\rho \sim const$ and the general solution to equation (1) --
harmonic oscillations with frequency $\omega_z$ -- has the
following form:
\begin{equation}
Z = Z_0 {\rm cos} (\omega_z t) + (V_{Z(0)}/\omega_z) {\rm sin}
(\omega_z t),
\end{equation}
where $Z_0$ and $V_Z(0)$ are the initial vertical coordinate and
vertical velocity component, respectively. It can be easily shown
that:
$$
(\sigma Z)^2 = (1/2)((\sigma V_{Z(0)})^2 + (P_Z\sigma
V_{Z(0)}/2\pi)^2) +
$$
\begin{equation}
(1/2) ((\sigma V_{Z(0)})^2 - (P_Z\sigma V_{Z(0)}/2\pi)^2) {\rm
cos} (4\pi/P_Z t)
\end{equation}
$$
+(1/2)(r P_Z\sigma Z_0 \sigma V_{Z(0)}/2\pi) {\rm sin}(4\pi/P_Z
t).
$$
It thus follows that by analyzing the dependence of {\bf
observed} $(\sigma Z)^2$ on age $t$ {\bf computed in terms of a
certain evolutionary grid}) we can infer four very interesting
parameters. First, the period $P_Z$ of vertical oscillations
(e.g., by power-spectrum analysis as is usually done in
variable-star research) and the {\bf local mass density $\rho_0$}
it implies via Poisson equation. Second and third, the initial
dispersion of {\bf vertical coordinates ($\sigma Z_0$)} and {\bf
vertical velocity components ($\sigma V_{Z(0)}$)}. And, fourth,
the {\bf correlation coefficient $r$} between the initial vertical
coordinate and vertical velocity. And, finally, by comparing the
local mass densities $\rho_0$ thus inferred with the local mass
density values based on recent analyses of stellar kinematics we
can discriminate between {\bf different evolutionary grids}
which, naturally, yield different individual cluster ages, and,
consequently, different periods $P_Z$.

\section{Initial data}

Our initial sample consisted of the catalog of 203 young open
clusters with heliocentric distances and ages determined in a
homogeneous way from published $UBV$ photoelectric or CCD
photometry (Dambis \cite{dambis}) using Kholopov's
(\cite{Kholopov}) empiric ZAMS and isochrones by Maeder \& Meynet
(\cite{MM}). Since one of the principal aims of this work is to
choose between stellar models with and without convective
overshooting, we first converted our cluster ages to two age
scales determined by the evolutionary grids of Pols (\cite{pols})
computed with and without overshooting.

\section{Evolution of vertical dispersion}
Figure~1 shows the variation of the 'sliding dispersion'
$\sigma_Z$ as a function of mean age $t$ for clusters located
within the Galactocentric distance interval from $R_0$ - 1 kpc to
$R_0$ + 1 kpc ($R_0$ = 7.5 kpc (see, e.g., Dambis \cite{dambis2}
and Dambis \& Rastorguev \cite{dr}) is the Galactocentric distance
of the Sun) computed in terms of the evolutionary grid of Pols et
al. (\cite{pols}) without overshooting. Figure~2 shows the
corresponding periodogram. Figures~3 and 4 show the corresponding
plots for the case where cluster ages are computed in terms of
the evolutionary grids of of Pols {\em et al.\/} (\cite{pols})
with overshooting. The inferred parameters $\sigma Z_0$, $\sigma
V_{Z(0)}$, $r(Z_0, V_{Z(0)}$, $P_Z$, and $\rho_0$ (computed
assuming that $A$ = +17.0~km/s/kpc and $B$ = -11.7~km/s/kpc -
Rastorguev {\em et al.\/} (\cite{rgdz}); note that adopting a
different pair of reasonable values of $A$ and $B$ based on
observations has little effect on the final result -- e.g.,
$\rho_0$ changes by only about 3\% if we adopt flat rotation
curve with $A+B$ = 0).

\subsection{Implications for stellar evolution}
Figure~5 compares our local mass density values computed in terms
of two evolutionary grids with recent estimates based on the
analysis of local stellar kinematics and the estimate of the
local density of visible mass (Holmberg \& Flynn \cite{hf}). It
is immediately evident from this figure that {\bf strong
overshooting should be ruled out} for stars in young clusters:
the local mass density $\rho_0 = 0.075 \pm 0.003 M_{odot}/pc^3$
implied by the corresponding grid is inconsistent not only with
the total (dynamic) mass density of $\rho_{dyn} = 0.102 \pm 0.010
M_{odot}/pc^3$ (Holmberg \& Flynn \cite{hf}), but even with the
density of visible matter $\rho_{dyn} = 0.095 M_{odot}/pc^3$
(ibid).

\subsection{Implications for local mass density}
Given that all known departures from standard models of stellar
evolution (e.g., mass loss, overshooting) slow down the rate of
evolution and, consequently increase the ages and times scales
based on these ages, the results summarized in Table~1 lead us to
conclude that the local mass density value inferred in terms of
evolutionary grids without overshooting should be considered as an
upper limit for this parameter. It thus follows that

$\rho_{max} = 0.118 \pm 0.006 M_{odot}/pc^3$.

\subsection{Implications for local dark mass}
Given the estimate of local density of visible mass ($\rho_{vis}
= 0.095 M_{odot}/pc^3$) mentioned above and the upper limit for
the total local mass density just obtained, we find that the local
density of dark mass cannot exceed
\begin{equation}
\rho{DM} \le \rho_{max} - \rho_{vis} = 0.027 M_{\odot}/pc^3.
\end{equation}

\subsection{Implications for star formation}
Two conclusions concerning star formation can be drawn from Figs~1
and 3 and Table~1. First, the population of open clusters at the
time of birth is overheated (i.e., has excess kinetic energy) in
the vertical direction: one can see that the layer of newly-born
clusters expands immediately after age zero. And second, there is
strong positive correlation (with a correlation coefficient of
0.8) between the initial vertical coordinates of newly-born
clusters and the initial vertical velocities. The two results
combined favor scenarios where star formation is triggered by
impacts of some massive bodies onto the Galactic plane.

\twocolumn

\begin{figure}
\includegraphics[width=5cm]{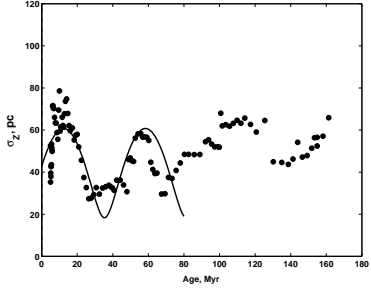}
\qquad \caption{Variation of the dispersion of vertical
coordinates ($\sigma_Z$) of young open clusters as a function of
age computed in terms of models of Pols {\em et al.\/}
(\cite{pols}) with overshooting.}
\end{figure}

\begin{figure}
\includegraphics[width=4.5cm]{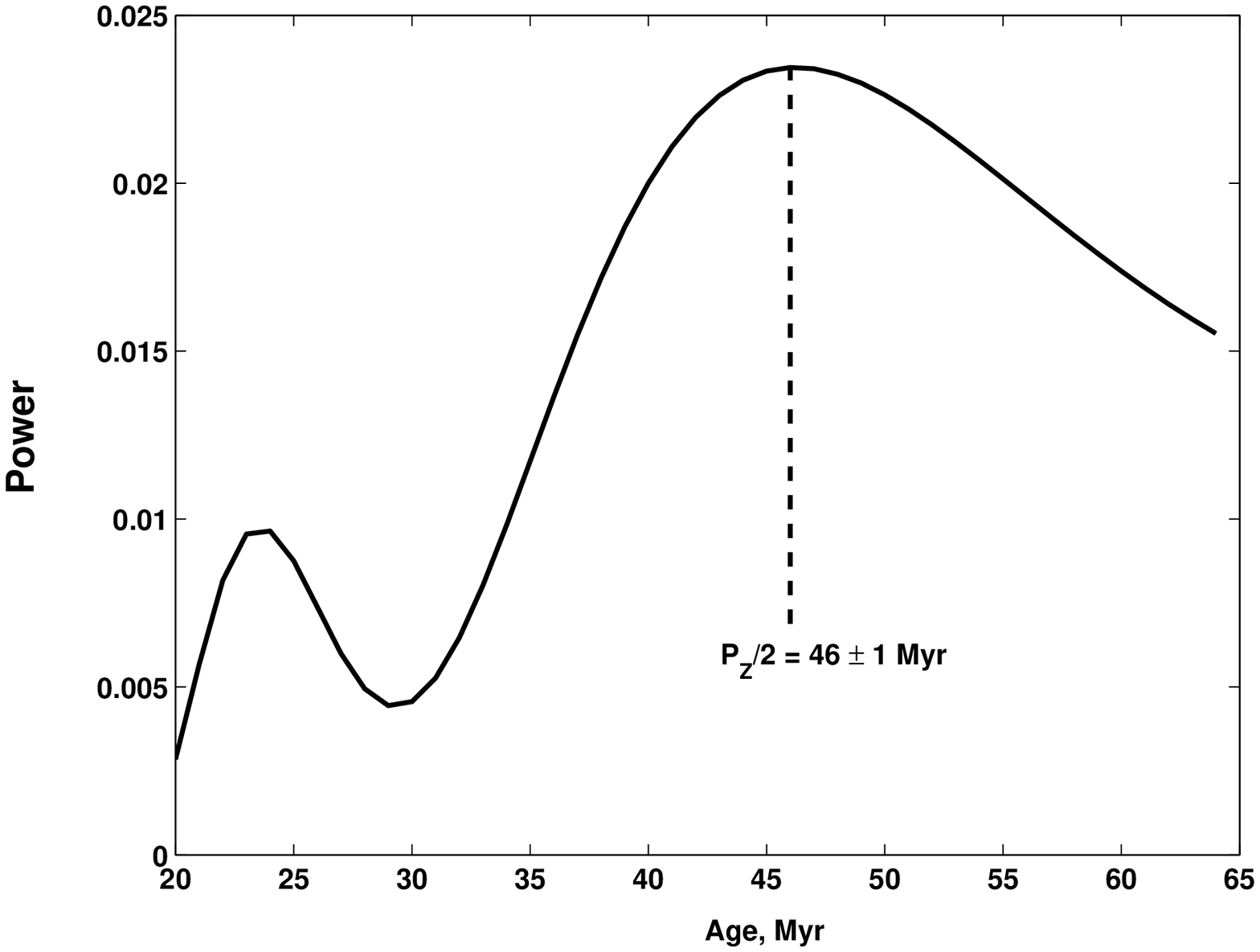}
\qquad \caption{Power spectrum of the variation of squared
vertical coordinate $z^2$ with cluster ages computed in terms of
models of Pols {\em et al.\/} (\cite{pols}) with overshooting.
$P_Z$ = 92 $\pm$ 2 Myr. }
\end{figure}

\begin{figure}
\includegraphics[width=4.5cm]{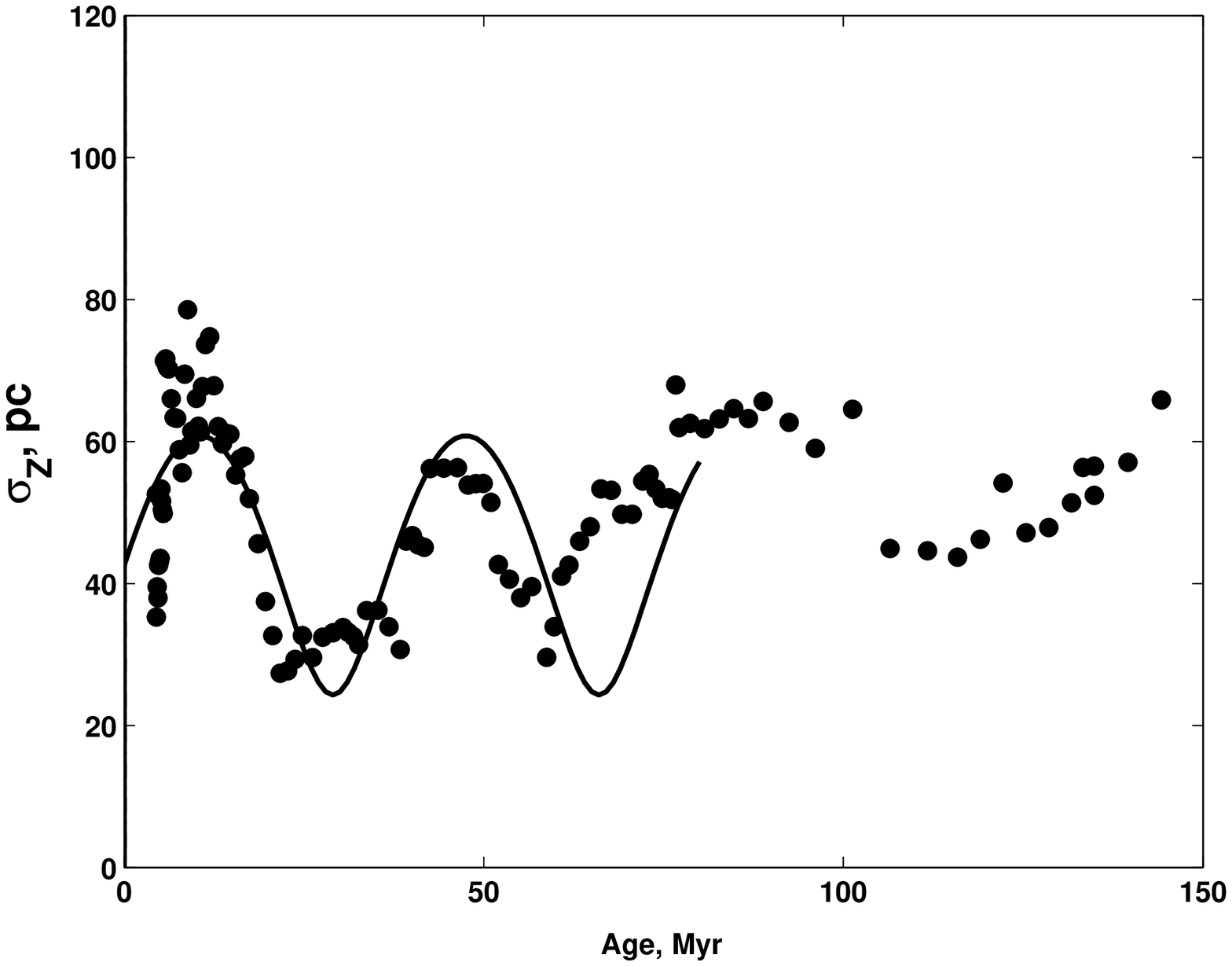}
\qquad \caption{Variation of the dispersion of vertical
coordinates ($\sigma_Z$) of young open clusters as a function of
age computed in terms of models of Pols {\em et al.\/}
(\cite{pols}) without overshooting.}
\end{figure}

\begin{figure}
\includegraphics[width=4.5cm]{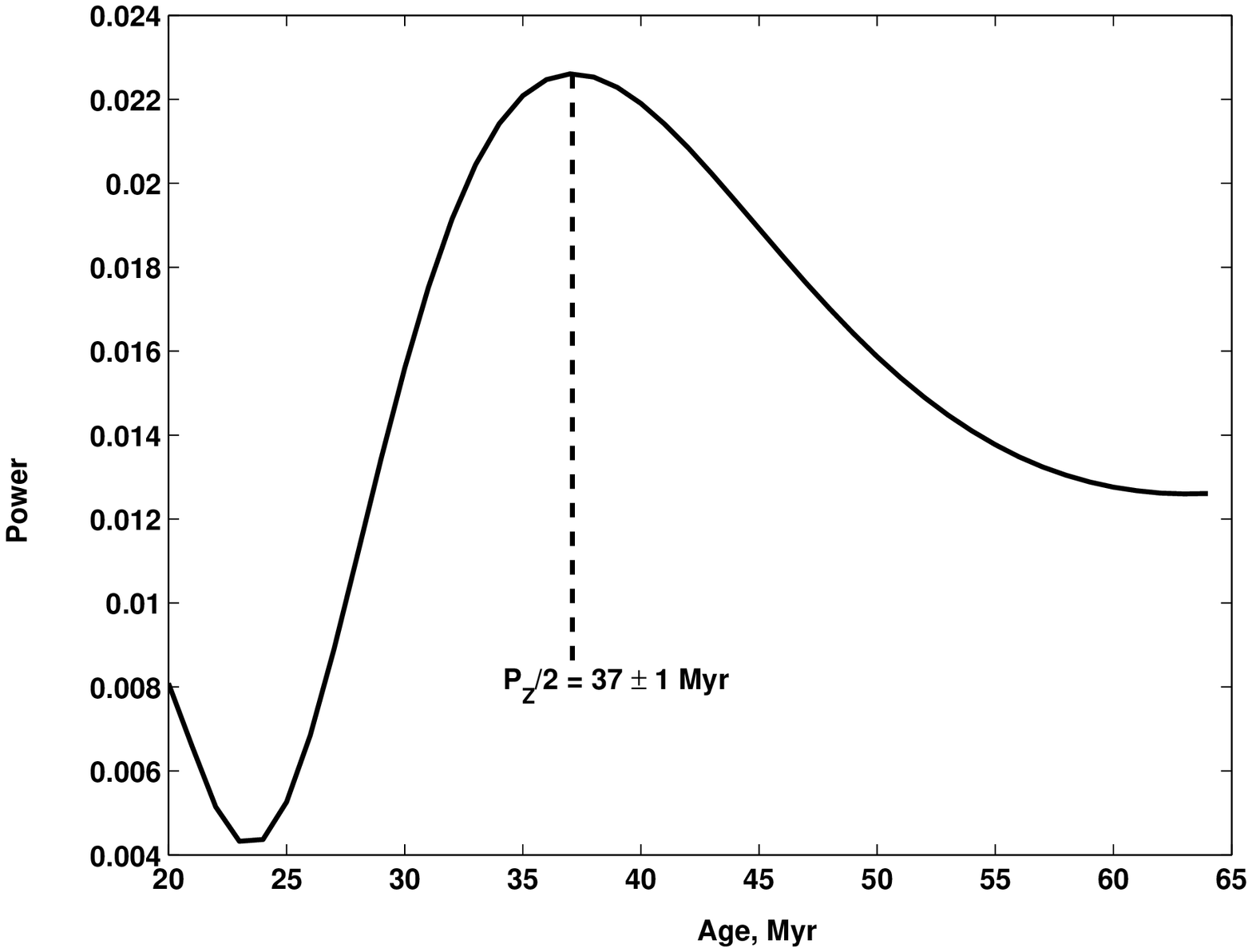}
\qquad \caption{Power spectrum of the variation of squared
vertical coordinate $z^2$ with cluster ages computed in terms of
models of Pols {\em et al.\/} (\cite{pols}) without overshooting.
$P_Z$ = 74 $\pm$ 2 Myr. }
\end{figure}

\begin{figure}
\includegraphics[width=4.5cm]{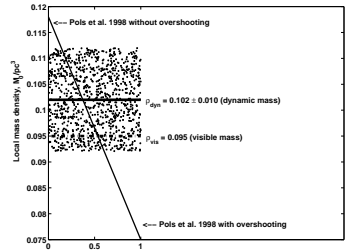}
\qquad \caption{Local mass density implied by the frequency of
oscillations of the dispersion of vertical cluster coordinates
($\sigma_Z$) as a function of "overshooting index".}
\end{figure}

\onecolumn

Another interpretation may involve vertical motions triggered by
with spiral density waves (see, e.g., Fridman {\em et al.\/}
(\cite{fridman}) who found vertical velocities in the spiral
galaxy NGC~3631 to be correlated with the arrangement of spiral
arms).

\begin{table}
  \centering
  \caption{Initial dispersions of  vertical coordinates ($\sigma Z_0$) and velocities
  ($\sigma V_{Z(0)}$) of young open clusters, the correlation coefficient $r$ between $Z_0$ and $V_{Z(0)})$,
  period $P_Z$ of vertical oscillations,
   and the implied local mass density
  (models of Pols {\em et al.\/} (\cite{pols}) without (STD) and with (OVS) overshooting).}\label{tab1}
\begin{tabular}{|r r r r r r|}
\hline Models & $\sigma Z_0$ & $\sigma V_{Z(0)}$ & $r$ & $P_Z$ & $\rho_0$ \\
    &  pc &  km/s &   &  Myr &  $M_{odot}/{\rm pc}^3$  \\ \hline
 STD & 40 & 4.3 & 0.81 & 74 & 0.118 \\
 & $\pm$ 2  & $\pm$ 0.4 & $\pm$ 0.08 & $\pm$
2 & $\pm$ 0.006 \\
&&&&& \\
OVS & 42 & 3.3 & 0.87 & 92 & 0.075 \\
 & $\pm$ 2  & $\pm$ 0.3 & $\pm$ 0.08 & $\pm$
2 & $\pm$ 0.003 \\
\hline
  \end{tabular}
\end{table}

\section{Acknowledgments}
The work was supported by the Russian Foundation for Basic
Research, grant nos. 01- 02-06012, 00-02-17804, 99-02-17842, and
01-02- 16086; Astronomy State Research and Technology Program;
and the Council for the Support of Leading Scientific Schools,
grant no. 00-15-96627.


\end{document}